\documentclass[12pt,superscriptaddress,showpacs,preprintnumbers,amsmath,showpacs]{revtex4}
\usepackage{graphicx}
\usepackage[subnum]{cases}

\def\be{\begin{equation}}
\def\ee{\end{equation}}
\def\ba{\begin{eqnarray}}
\def\ea{\end{eqnarray}}
\def\bw{\begin{widetext}}
\def\ew{\end{widetext}}

\begin{document}

\title{Wannier-Stark ladders in one-dimensional elastic systems}

\author{L. Guti\'errez}
\affiliation{Centro de Ciencias F\'isicas, Universidad Nacional
Aut\'onoma de M\'exico, P.O. Box 48-3, 62251 Cuernavaca Mor., M\'exico}

\author{A.~\surname{D\'{\i}az-de-Anda}}
\affiliation{Centro de Ciencias F\'isicas, Universidad Nacional
Aut\'onoma de M\'exico, P.O. Box 48-3, 62251 Cuernavaca Mor., M\'exico}

\author{J. Flores}
\altaffiliation{Permanent address: Instituto de F\'{\i}sica, 
Universidad Nacional Aut\'onoma de M\'exico, P.O. Box 20-364, 
01000 M\'exico, D. F., M\'exico}
\affiliation{Centro de Ciencias F\'isicas, Universidad Nacional
Aut\'onoma de M\'exico, P.O. Box 48-3, 62251 Cuernavaca Mor., M\'exico}

\author{R.~A.~\surname{M\'endez-S\'anchez}}
\affiliation{Centro de Ciencias F\'isicas, Universidad Nacional
Aut\'onoma de M\'exico, P.O. Box 48-3, 62251 Cuernavaca Mor., M\'exico}

\author{G. Monsivais}
\affiliation{Instituto de F\'{\i}sica, Universidad Nacional 
Aut\'onoma de M\'exico, P.O. Box 20-364, 
01000 M\'exico, D. F., M\'exico}

\author{A. Morales}
\affiliation{Centro de Ciencias F\'isicas, Universidad Nacional
Aut\'onoma de M\'exico, P.O. Box 48-3, 62251 Cuernavaca Mor., M\'exico}

\begin{abstract}
The optical analogues of Bloch oscillations and their associated Wannier-Stark 
ladders have been recently analyzed. In this paper we propose an elastic 
realization of these ladders, employing for this purpose the torsional vibrations 
of specially designed one-dimensional elastic systems. We have measured, for the first time, 
the ladder wave amplitudes, which are not directly accessible either in the quantum mechanical 
or optical cases. The wave amplitudes are spatially localized and coincide rather well 
with theoretically predicted amplitudes. The rods we analyze can be used to localize 
different frequencies in different parts of the elastic systems and viceversa.
\end{abstract}


\maketitle

Recently, undulatory systems showing analogues of Bloch oscillations and 
Wannier-Stark ladders (WSL) attracted increasing attention in several 
fields of 
physics~\cite{Agarwaletal,Morandottietal,Sapienzaetal,Ghulinyanetal,Lanzillottietal}.
As shown by Bloch, electrons in a periodic potential have extended solutions. 
The same is true for the behavior of an electron under the action 
of a static electric field. In contrast, and opposite to intuition, when both 
the periodic potential and the electric field are present, the solutions are 
localized; this is only true when band to band Zener tunneling is negligible 
or the system is short enough. The spectrum then shows equally spaced 
resonances known as Wannier-Stark ladders, the nearest--neighbor level 
spacing being proportional to the intensity of the external field ~\cite{Wannier}. 
In the time domain, the Wannier-Stark ladders yield the so  
called Bloch oscillations which consist in a counterintuitive effect where the 
electrons show an oscillatory movement under the action of the static 
external electric field~\cite{Bloch,Zener}.

The predictions by Bloch and Wannier gave rise to a 
long controversy that lasted more than 60 years for the Bloch 
oscillations \cite{HartEmin}, and more than 20 years for the 
WSL \cite{FukuyamaBariFrogedby,Rabinovitch,BanavarCoon}. 
The ladders 
were observed before Bloch oscillations. This was done first in 
numerical experiments using simple one-dimensional models 
\cite{NagaiKondo}
and later in the laboratory \cite{MendezArgullo-RuedaHong}.
Bloch oscillations were also observed later on \cite{Feldmannetal}. 
The most important ingredient to explain WSL is the wavelike 
behavior of the electrons. Therefore, these ladders could also 
be observed in classical undulatory systems. 
Some of these classical systems have been analyzed theoretically 
\cite{MonsivaisCastillo-MussotClaro,MateosMonsivais,Monsivaisetal}.

\begin{figure}[t]
\includegraphics[width=\columnwidth]{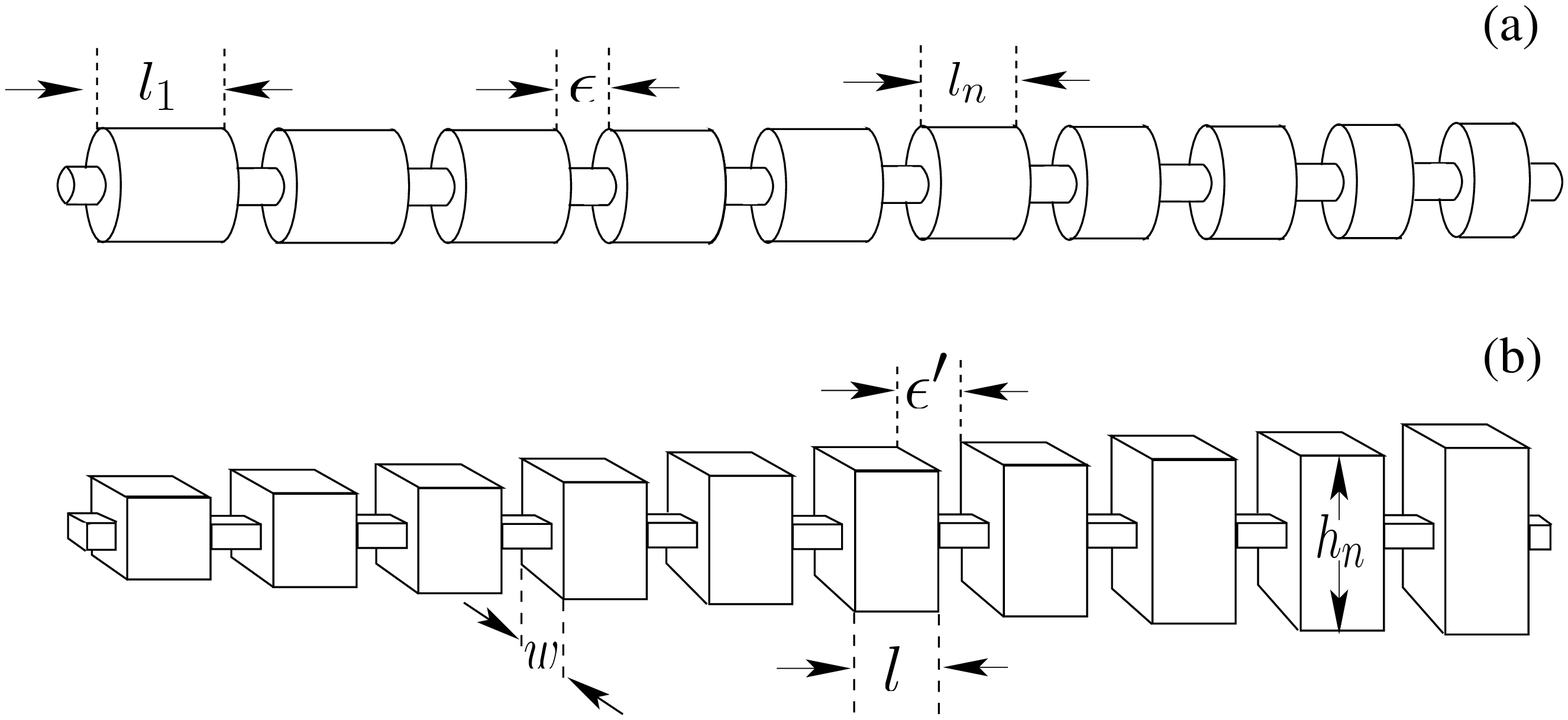}
\caption{Rods used to obtain the Wannier-Stark ladders: (a) 
rod with varying length cells and (b) beam with 
equal--length cells but different heights. 
For system A, $l_n=l/(1+n\gamma)$, $n=1,\dots,14$, with 
$l=10.8$~cm, $\gamma=0.091$, $\epsilon=2.52$~mm and 
$\sqrt{G/\rho}=3104.7$~m/s. 
The radii of the small and big cylinders are $r=2.415$~mm 
and $R=6.425$~mm, respectively. 
In system B, $l=5.0$~cm, $w=1.905$~cm and $c_n=c(1+n \gamma)$,
$n=1,\dots,15$ with $c=2027.3$~m/s and $\gamma=0.02786$. The width, height 
and length of the small cuboids are $w'=5.0$~mm, $h'=5.0$~mm, 
and $\epsilon'=6.0$~mm, respectively.}
\label{Fig:Rods}
\end{figure}

In this paper we study special elastic rods whose torsional waves 
for free ends present some analogies to the WSL. The first system, depicted 
in Fig.~\ref{Fig:Rods} (a) and which will be referred to as system A, 
consists of a set of $N$ circular cylinders of radius $R$ and varying 
length $l_n$, $n=1,2,\dots,N$, separated by very small cylinders 
of length $\epsilon \ll l_n$ and radius $r<R \ll l_n$.
This is the elastic analogue of the optical 
system with varying widths used in Ref.~\cite{Agarwaletal}. System B, shown in 
Fig.~\ref{Fig:Rods} (b), is a beam formed by $N$ cuboids 
of constant width $w$ and length $l$. They have different 
heights $h_n$, for $n=1,2,\dots,N$, with $w, h_n \ll l$. 
These cuboids are separated by small cuboids of 
dimensions $h'=w',\epsilon'\ll l$.
This is the elastic analogue of the optical systems with a gradient 
of the refractive index along the direction of propagation used in 
Refs.~\cite{Morandottietal,Sapienzaetal,Ghulinyanetal}, 
as we shall see below. Systems A and B were constructed by 
machining a solid aluminum piece.

We first discuss the design of these systems and then, 
from a qualitative point of view, the normal--mode 
frequencies and wave amplitudes for torsional vibrations.
We later use the transfer matrix method and obtain the 
normal--mode properties, which will be compared to 
experimental measurements. 

In order to design systems A and B we start with 
what could be called an independent rod model in which 
each body oscillates independently from the rest. The 
normal--mode torsional frequencies $f_j^{(n)}$ of rod $n$ 
with length $l_n$ and wave velocity $c_n$ are given by 
the well-known expression 
\cite{Graff}
\be
f_j^{(n)}=\frac{c_n}{2 l_n} j,
\label{Eq:FrequenciesAnyRod}
\ee
where $j$ is the number of nodes in the wave amplitude. 
To get a set of equidistant frequencies we consider two options: 
either the lengths $l_n$ are varied with a fixed wave 
velocity or the lengths are kept constant and the wave velocity is 
changed. 
For system A we take circular rods with $l_n=l/(1+n\gamma)$, 
$n=1,\dots,N$ and $c_n=(G/\rho)^{1/2}$ were $\rho$ is the density, 
$G$ the shear modulus and $l$ a fixed arbitrary length.
Notice that, in circular rods, the velocity does not depend on 
the radius \cite{Graff}.
For system B we take $l_n=l$ and $c_n  = c(1 + n \gamma)$, 
$c$ being an arbitrary constant velocity. The parameter 
$\gamma$ is dimensionless.

To construct system B we use the Navier expression for the torsional 
velocity of cuboid $n$: 
\be
c_n=\sqrt{\frac{G \alpha_n}{\rho I_n}}
\label{Eq:VelocityB}
\ee
where $I_n = (h_n w^3+h_n^3 w)/12$ is the moment of inertia with respect 
to the axis of the system and $\alpha_n$ is given by 
\ba
\alpha_n&=&\frac{256}{\pi^6}\sum_{m=0}^{\infty}\sum_{p=0}^{\infty}
\frac{1}{\left(2m+1\right)^2\left(2p+1\right)^2} \times \nonumber \\
&&\frac{h_n w}{\left(({2m+1})/{h_n}\right)^2
+\left(\left(2p+1\right)/w\right)^2}.
\label{Eq:Alphas}
\ea
For beams of varying $h_n$ we have verified Eqs.~(\ref{Eq:VelocityB}) 
and~(\ref{Eq:Alphas}) experimentally. 
Solving these equations 
we have obtained the values of $h_n$  such that 
$c_n = c(1 + n \gamma)$. 

Then
\begin{numcases}{f^{(n)}_j=}
\label{Eq:FrequenciesA} \sqrt{{G}/{\rho}} (1+n\gamma) j/2 l 
& \textrm{for system A} \\
\label{Eq:FrequenciesB} c (1 + n \gamma) j/2 l & \textrm{for system B}, 
\end{numcases}
and the differences $\Delta f^{(n)}_j=f^{(n+1)}_j-f^{(n)}_j$ are equal to
\begin{numcases}{\Delta_j\equiv \Delta f^{(n)}_j=}
\label{Eq:DeltaA} \sqrt{{G}/{\rho}}\gamma j/2 l& \textrm{for system A}\\
\label{Eq:DeltaB} c \gamma j/{2 l} & \textrm{for system B},
\end{numcases}
which are independent of index $n$. 

We shall now discuss system A.
When the arbitrary parameter $\gamma$ is set equal to zero, 
a locally periodic rod is formed. This locally periodic rod 
shows a band spectrum \cite{Moralesetal}. 
When $\gamma\ne0$ a completely different 
spectrum occurs. The new spectrum then resembles the 
Wannier--Stark ladder.

Before presenting the calculation of the normal modes for this system, 
and then showing numerical and experimental results, let us make a 
qualitative analysis to see what type of spectrum could be expected 
from the independent rod model. 
At the lowest frequencies, the wavelength $\lambda$ is of the same 
order of 
magnitude as $L\approx \sum_{n=1}^N l_n$, and the whole rod is 
excited. But when $\lambda$ decreases and 
becomes of the order of  $l_1=l/(1+\gamma)$, the longest rod, 
say rod 1, is excited in 
a state equivalent to its lowest normal mode. The rest of the $N$ rods 
are out of resonance, so the amplitude decreases as we move farther away 
from rod 1. Therefore, the state is localized around the latter. In some 
sense this was to be expected since we are disturbing a periodic structure 
to obtain a disordered one--dimensional system, which always 
shows localized wave amplitudes. Increasing 
the exciting frequency by $\Delta_1$ the rod with length 
$l_2=l/(1+2\gamma)$, that is, rod 2, will now be excited and the 
rest will be out of resonance.  The amplitudes of the vibrations 
therefore decrease as their distance from rod 2 
increases. The wave amplitude is again localized but now around rod 2; it 
has a similar shape as the wave amplitude that rod 1 had before, 
but it has been slightly deformed, squeezed, and translated from 
rod 1 to rod 2. The same argument applies when rod $n$ of 
length $l_n=l/(1+n\gamma)$ is excited. 

What we have done is to produce a finite WSL, i.e., $N$ localized 
states with constant difference in frequency given by Eq.~(\ref{Eq:DeltaA}). 
However, more ladders exist since
normal modes with two or more nodes can also be excited in each rod. 
For instance, taking $j=2$ in Eq.~(\ref{Eq:FrequenciesA}) a second 
ladder is obtained.  
This ladder is different from the first one because the 
frequency difference 
is now twice the one of the lower WSL, as can be seen from 
Eq.~(\ref{Eq:DeltaA}). 
The states are again localized and all have a similar shape although 
squeezed. A third ladder exists with $\Delta_3=3 \Delta_1$ 
and so on for other values of $j$. A similar 
argument for system B shows also the existence of several WSL. 
The difference between the quantum-mechanical and elastic ladders 
is that in the latter the spacing between resonances is not the same  
for different ladders. 

\begin{figure}[t]
\includegraphics[width=\columnwidth]{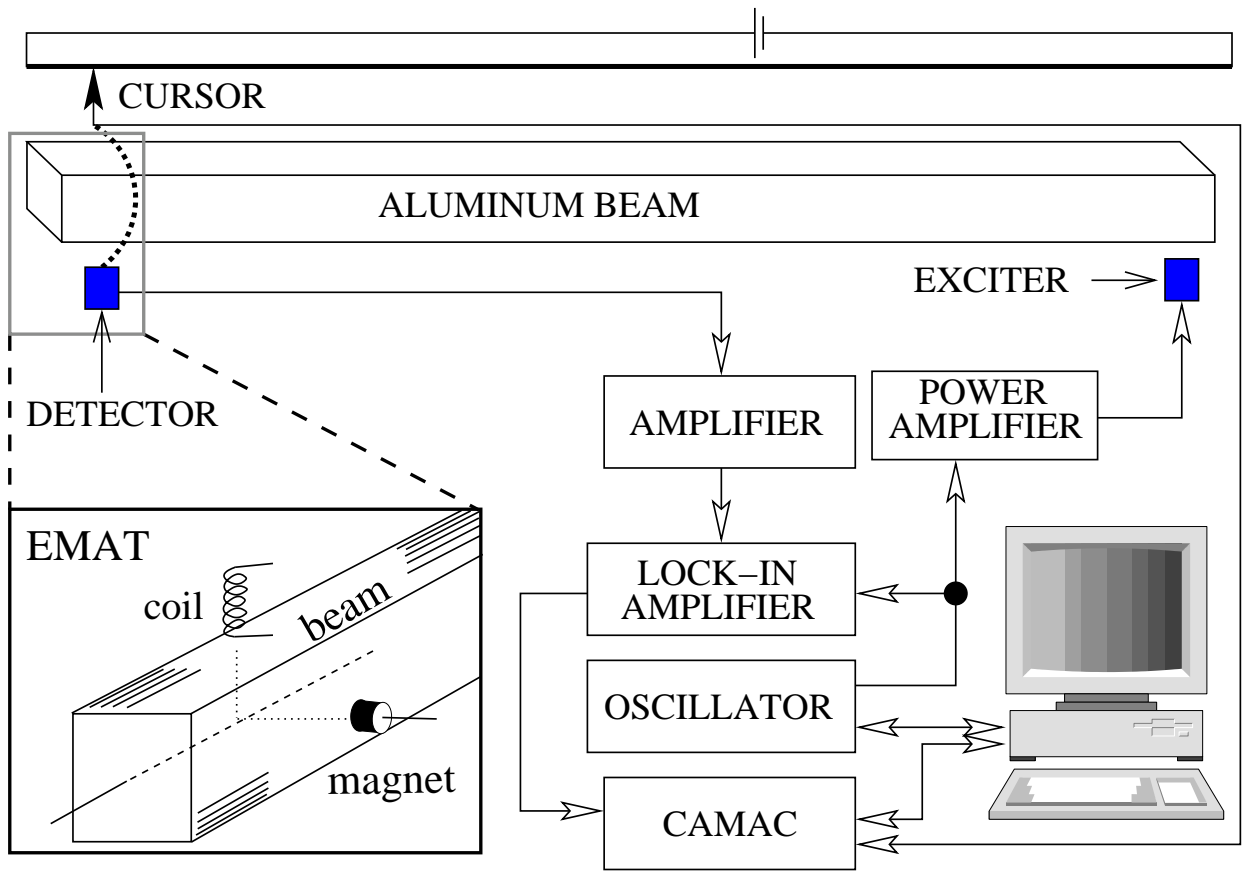}
\caption{Block diagram of the experimental setup. For system A 
both detector and exciter are EMATs, while for system B the 
exciter was a piezoelectric transducer.}
\label{Fig:BlockDiagram}
\end{figure}

We have calculated the eigenmode properties of the rods of 
Fig.~\ref{Fig:Rods}~(a) and~(b) with free--end boundary conditions
using the transfer matrix method for torsional waves discussed 
in Ref.~\cite{Moralesetal}. 
The normal mode frequencies and amplitudes were measured using 
the experimental set up described in Fig.~\ref{Fig:BlockDiagram}. 
We use an electromagnetic acoustic transducer (EMAT) 
which is very versatile and operates at low 
frequencies. This EMAT, which we have recently developed 
\cite{MoralesGutierrezFlores}, can selectively excite 
or detect compressional, torsional or flexural vibrations. In the inset of Fig.~\ref{Fig:BlockDiagram} the EMAT has been installed 
to detect torsional vibrations; see Ref.~\cite{Moralesetal}. 

\begin{figure}[t]
\includegraphics[width=\columnwidth]{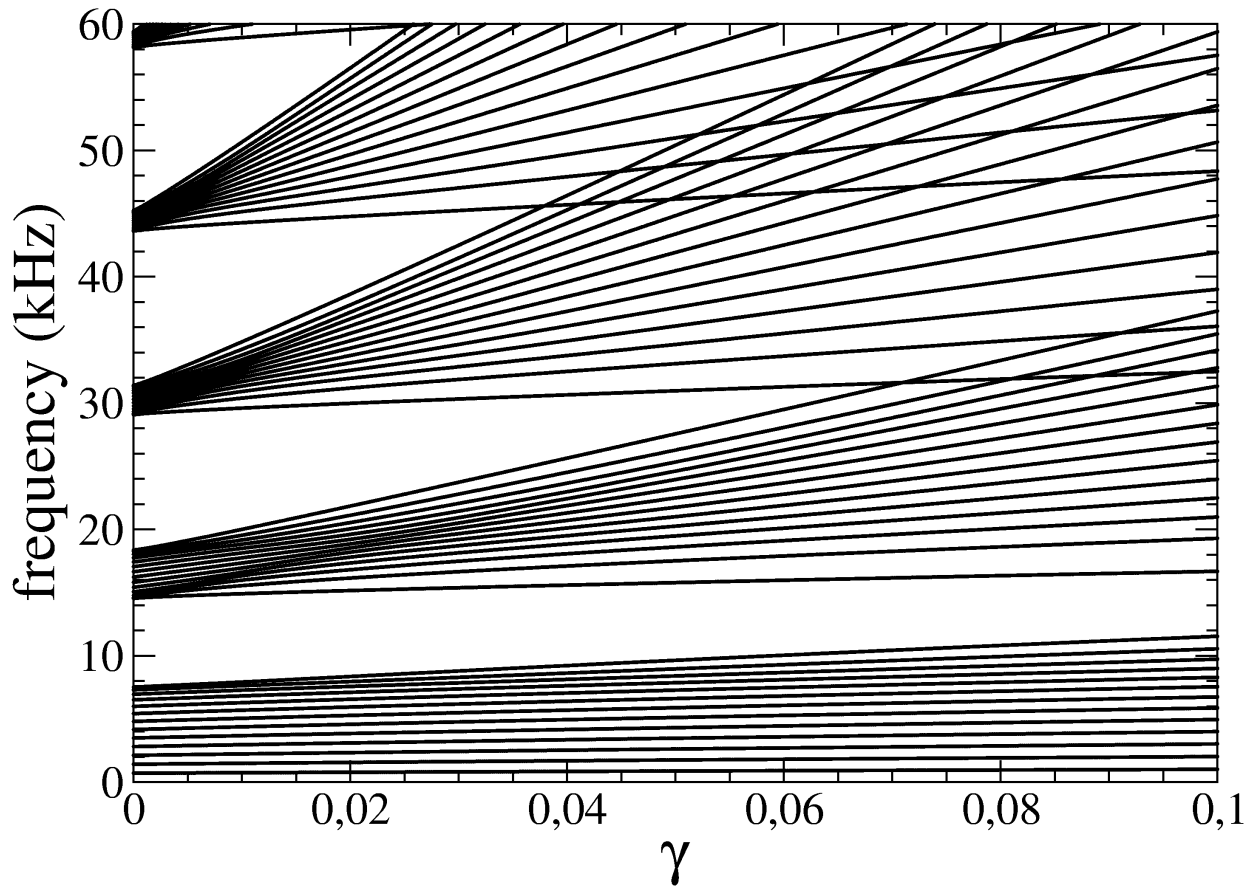}
\caption{Normal--mode frequencies of System A as a function of the 
dimensionless parameter $\gamma$. 
}
\label{Fig:Evolution}
\end{figure}

\begin{figure}[t]
\includegraphics[width=\columnwidth]{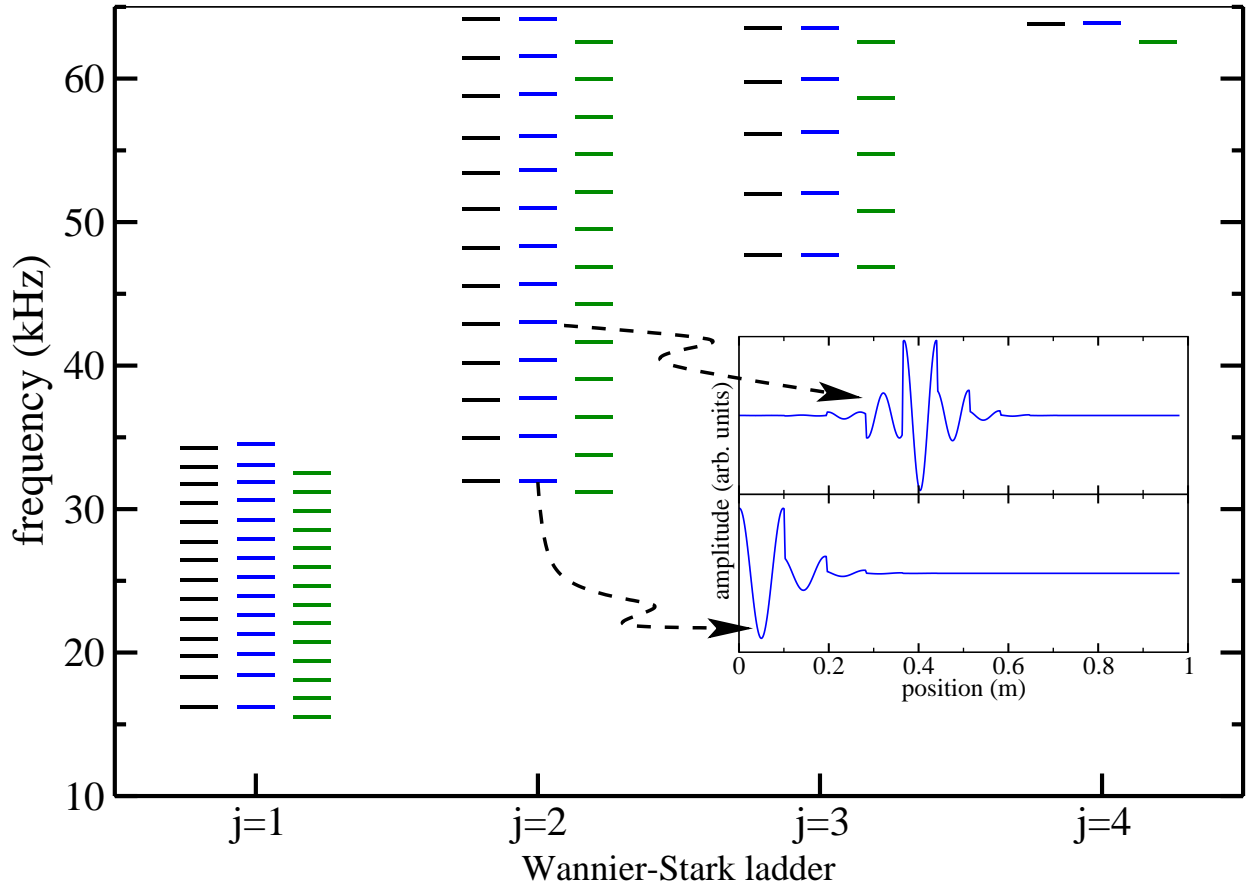}
\caption{(color online) Normal--mode frequencies of System A 
yielding the elastic 
Wannier--Stark effect. For each value of $j$ the left hand side column 
corresponds to the experimental values, the middle column to the 
numerical results obtained using the transfer matrix method and the 
right hand side shows the approximate results following from the 
independent rod model. 
In the calculation we used an effective value for $r/R$ \cite{Moralesetal}.
The uncertainty in the experimental values is less than 0.01\%.
In the insets the theoretical wave amplitudes are given for 
a state at the extreme of the ladder and another one in the center of it. 
}
\label{Fig:Frequencies}
\end{figure}

\begin{figure}[t]
\includegraphics[width=\columnwidth]{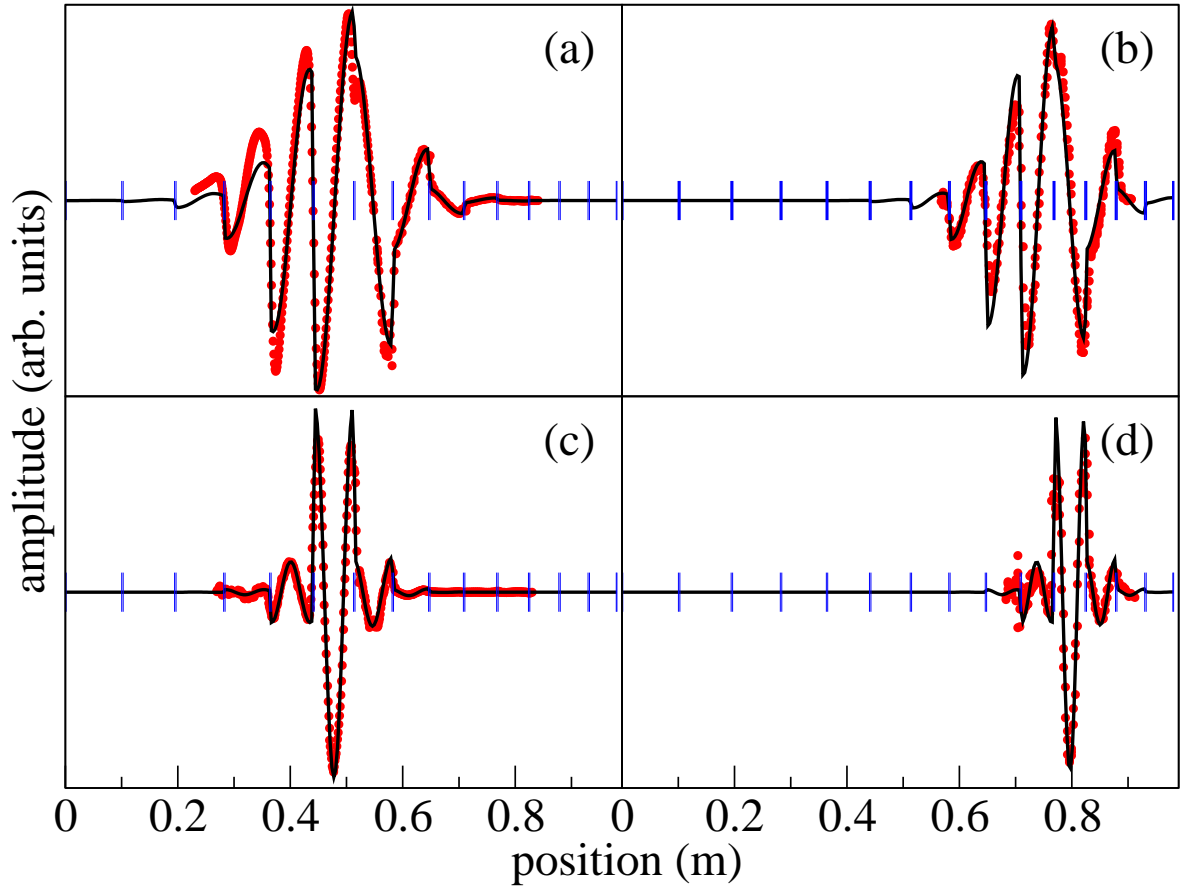}
\caption{(color online) Wave amplitudes of system A for 
19 nodes (a), 23 nodes (b), 
33 nodes (c), and 41 nodes (d). The continuous line corresponds to the 
transfer-matrix results and the dots to the measurements. 
The vertical lines along the rod axis indicate the position of 
the notches.}
\label{Fig:WaveAmplitudesSystemA}
\end{figure}

We show in 
Fig.~\ref{Fig:Evolution} the spectrum of system A 
as a function of the dimensionless parameter $\gamma$.
As mentioned above, for $\gamma = 0$ a band spectrum 
appears, and as $\gamma$ grows the levels of each band 
separate to form the WSL.
The normal--mode frequencies of the rod shown in Fig.~\ref{Fig:Rods} (a) 
are given in Fig.~\ref{Fig:Frequencies} 
for  $\gamma = 0.091$. We first note that the theoretical 
results coincide very well with the experimental ones. Furthermore, 
the qualitative treatment provides a rather good first approximation.
One can see from this figure that the 
states form a set of Wannier--Stark ladders, as discussed before. 
The first band composed by the extended modes 
is not displayed in this graph. 
Notice that the frequencies in the extremes of each ladder 
do not have 
the same difference in frequency as those at the middle of the ladder.
This is due to a border effect in the wave amplitudes localized near 
the free ends. As shown in the insets of Fig.~\ref{Fig:Frequencies} 
the border amplitude lacks a portion of the wave amplitude that the 
states at the center of the ladder have.

In Fig.~\ref{Fig:WaveAmplitudesSystemA} we show the comparison 
of theoretical and experimental wave 
amplitudes. These are localized around rod $n$.  
For example, in Fig.~\ref{Fig:WaveAmplitudesSystemA}~(a) the 
sixth rod resonates and in 
Fig.~\ref{Fig:WaveAmplitudesSystemA}~(b) another state corresponding 
to the same ladder is localized around the tenth rod. 
Both have the same form although squeezed. 
Figures~\ref{Fig:WaveAmplitudesSystemA}~(c) 
and~\ref{Fig:WaveAmplitudesSystemA}~(d) show two states of the 
second ladder, with $n = 6$ and $n = 11$, respectively. 
Localization is again observed and, as was 
to be expected, the amplitudes now show two 
nodes in the resonating rods. Note the excellent agreement 
between theory and experiment after adjusting the 
height of the theoretical wave amplitude at only one point.

\begin{figure}[t]
\includegraphics[width=\columnwidth]{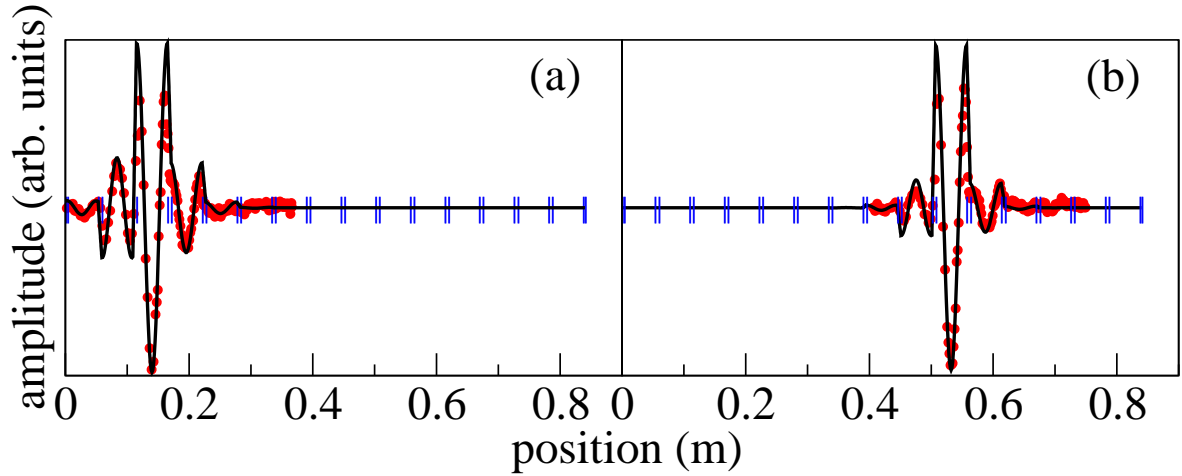}
\caption{(color online) Two wave amplitudes of system B, (a) one localized 
on the third rod with frequency $f=44.256$~kHz and (b) other 
localized on the tenth rod with frequency $f=51.258$~kHz. 
The double vertical lines along the beam axis indicate 
the position of the notches.}
\label{Fig:WaveAmplitudesSystemB}
\end{figure}

As mentioned above, system B shows similar properties. 
We present in Fig.~\ref{Fig:WaveAmplitudesSystemB}, as 
an example, two wave amplitudes for the first WSL of system B. 
In contrast with system A, these wave amplitudes have the 
same shape but translated and are not squeezed.
Notice that the one--dimensional transfer matrix method 
fit the experimental wave amplitudes in spite of the fact that 
$w$ and $h_n$ are not much smaller than $l$. 

To conclude, we have constructed an elastic analogue 
of Wannier--Stark ladders. In contrast with the optical analogue of 
Refs.~\cite{Agarwaletal,Morandottietal,Sapienzaetal,Ghulinyanetal}
we have observed the WSL directly. Furthermore, we measured 
for the first time the wave amplitudes, including phases, which show 
localization.  We also observed higher Wannier--Stark ladders. 
The elastic Wannier-Stark ladders have potential applications in the design of 
systems with localized vibrations.

This work was supported by DGAPA-UNAM under projects 
IN104400 and IN104903-3 and by CONACyT M\'exico, project 
41024-F.

\end{document}